\documentclass[10pt]{article}    
\usepackage{amssymb} 
\usepackage{amsmath}  
\usepackage{algorithm}
\usepackage{algorithmic}
\usepackage{placeins}
\usepackage{array}
\usepackage{caption}
\usepackage{color}
\usepackage{tikz}
\usetikzlibrary{plotmarks}
\usepackage{pgfplots}
\usepackage{mathtools}
\usepackage{authblk}
\usepackage{lineno} 
\pgfplotsset{compat=1.17}

\newenvironment{customlegend}[1][]{%
	\begingroup
	\csname pgfplots@init@cleared@structures\endcsname
	\pgfplotsset{#1}%
}{%
	\csname pgfplots@createlegend\endcsname
	\endgroup
}
\def\addlegendimage{\csname pgfplots@addlegendimage\endcsname}
\pgfkeys{/pgfplots/number in legend/.style={%
		/pgfplots/legend image code/.code={%
			\node at (0.295,-0.0225){#1};
		},%
	},
}
\def\norm#1{\left\|#1\right\|}
\newcolumntype{C}[1]{>{\centering\arraybackslash\hspace{0pt}}b{#1}}

\newenvironment{keywords}
{\begin{trivlist}\item[]{\bfseries Keywords:}\ }
	{\end{trivlist}}

\newenvironment{AMS}
{\begin{trivlist}\item[]{\bfseries AMS:}\ }
	{\end{trivlist}}

\begin{document}
\title{An augmented wavelet reconstructor for\\atmospheric tomography}
\author[1]{Ronny Ramlau}
\author[1]{Bernadett Stadler}
\affil[1]{Industrial Mathematics Institute, Johannes Kepler University Linz, Altenbergerstra\ss e 69, 4040 Linz, Austria}

\maketitle

\begin{abstract}
	Atmospheric tomography, i.e. the reconstruction of the turbulence profile in the atmosphere, is a challenging task for adaptive optics (AO) systems of the next generation of extremely large telescopes. Within the community of AO the first choice solver is the so called Matrix Vector Multiplication (MVM), which directly applies the (regularized) generalized inverse of the system operator to the data. For small telescopes this approach is feasible, however, for larger systems such as the European Extremely Large Telescope (ELT), the atmospheric tomography problem is considerably more complex and the computational efficiency becomes an issue. Iterative methods, such as the Finite Element Wavelet Hybrid Algorithm (FEWHA), are a promising alternative. FEWHA is a wavelet based reconstructor that uses the well-known iterative preconditioned conjugate gradient (PCG) method as a solver. The number of floating point operations and memory usage are decreased significantly by using a matrix-free representation of the forward operator. A crucial indicator for the real-time performance are the number of PCG iterations. In this paper, we propose an augmented version of FEWHA, where the number of iterations is decreased by $50\%$ using a Krylov subspace recycling technique. We demonstrate that a parallel implementation of augmented FEWHA allows the fulfilment of the real-time requirements of the ELT.
\end{abstract}



\section{Introduction}
The image quality of the new generation of earthbound Extremely Large Telescopes (ELT) is heavily degraded by atmospheric turbulences triggered by the irregular mixing of cold and hot air. Due to these irregularities the refractive index of air becomes inhomogeneous, and thus the wavefronts arriving at the telescope pupil are distorted. These optical distortions can be compensated by using a technique called Adaptive Optics (AO) \cite{Roddier,RoWe96,ElVo09}, in which the deformations of the wavefronts of natural or laser guide stars (NGS or LGS) are measured with wavefront sensors (WFS) and, subsequently, corrected using deformable mirrors (DMs). Such a DM typically consists of a thin surface to reflect light and a set of actuators that drive the shape of the mirror. Several AO systems require the reconstruction of the turbulence profile in the atmosphere, which is called atmospheric tomography. These three AO systems are: Laser Tomography (LTAO), Multi Object Adaptive Object (MOAO) and Multi Conjugated Adaptive Optics (MCAO). LTAO uses several LGS and NGS with one DM to sharpen a single object of interest. MOAO is based on the same concept, but uses different mirrors that sharpen various objects of interest at the same time. MCAO uses several DMs, conjugated to different heights, and aims to achieve a high imaging quality in a large field of view. Detailed information about the systems can be found in \cite{Hammer02,Andersen06,RiElFl00,Puech08,DBB10}.

Mathematically, the atmospheric tomography problem is ill-posed, i.e., the relation between measurements and the solution is unstable \cite{Davison83,Nat86}. As a consequence, regularization techniques are applied. A common way to deal with this inverse problem is the Bayesian framework, because it allows to incorporate statistical information about turbulence and noise. Typically, the random variables are assumed to be Gaussian and the maximum a posteriori estimate (MAP) is used for an optimal point estimate of the solution. The dimension of the atmospheric tomography problem depends on the number of subapertures of the WFS and on the number of degrees of freedom of the DMs, which increased drastically during the last years. Since the atmosphere is changing rapidly, the solution has to be computed in real-time, i.e., within approximately $2~ms$. There are various ways on how to solve the atmospheric tomography problem, either directly or iteratively, see \cite{Fusco,ElGiVo02,GiElVo02b,GiElVo03,YaVoEl06,GiElVo07,GiEl08,RoCoGrScFu10,ThiTa10,Tallon_et_al_10,RaRo12,RoRa13,RaObRoSa13,SaRa15,RafRaYu16}. So far, the standard solver is the Matrix Vector Multiplication (MVM), i.e., the direct application of a (regularized) generalized inverse of the system operator. This direct solution method is suitable for small telescopes, however, for extremely large telescopes the computational efficiency becomes an issue. A promising alternative are iterative methods as, e.g., the Finite Element Wavelet Hybrid Algorithm. FEWHA utilizes a dual domain discretization approach in which the operators are transformed into a finite element or wavelet domain, leading to sparse representations of the underlying matrices. The dual domain discretization of the MAP estimate is then solved using a preconditioned conjugate gradient method (PCG). As already illustrated in \cite{stadler2020}, FEWHA is not perfectly parallelizable and for ELT-sized test configurations the real-time requirements are hard to fulfil on an off the shelf hardware system. A crucial indicator for the computational efficiency is the number of PCG iterations. Our novel method called augmented FEWHA speeds up the convergence of the PCG method by recycling information from previous time steps. In particular, the Krylov subspace, generated when solving the previous systems, is recycled by utilizing orthogonal projections in subsequent systems. The augmented conjugate gradient method for solving consecutive linear systems was proposed in \cite{AugCG}. The concept there is based on the idea of Saad in \cite{Saad87} on Krylov subspace recycling for solving linear systems with several right-hand sides. However, the work of Saad is based on the Arnoldi or Lanczos process. For a survey on subspace recycling techniques for iterative methods we refer to \cite{soodhalter2020survey}.

The paper is organized as follows: we start with a rough overview on the atmospheric tomography problem and the Bayesian framework for regularization in Section \ref{sec:atmo}. Afterwards, we recall the dual domain discretization approach of FEWHA in Section \ref{sec:discretization}. In Section \ref{sec:pcg} we present the well-known PCG method and an extension based on Krylov subspace recycling, called augmented PCG. In Section \ref{sec:augFEWHA} we propose the augmented FEWHA and demonstrate its quality and robustness by numerical simulations for ELT-sized test configurations in Section \ref{sec:numerics}. An analysis of the algorithm regarding computational performance, i.e., floating point operations, memory usage and runtime on CPU, is provided in Section \ref{sec:performance}.

\section{Atmospheric tomography}\label{sec:atmo}
In atmospheric tomography commonly a layered system of the atmosphere is assumed, with the aim to reconstruct the refractive index of the turbulent layers from WFS measurements \cite{RoWe96}.

The relation between the $L$ turbulent layers $\phi = (\phi_1, ..., \phi_L)$ and the wavefront sensor measurements $s$ in a guide star direction $g$ is given via the tomography operator $A$ by
\begin{equation}\label{eq:atmo}
s = (s_g)_{g=1}^G = A\phi.
\end{equation}
We consider here the Shack-Hartmann (SH) WFS, i.e., the average slope of the wavefront over the area of the lens, called subaperture, is determined by the vertical and horizontal shift of the focal points of the subapertures. The tomography operator is composed into a SH operator $\Gamma$ and a geometric propagation operator $P$ in the direction of a certain guide star $g$
\begin{equation*}
s_g = \Gamma_g P_g \phi \text{ for } g=1,...,G.
\end{equation*}

The SH operator $\Gamma$ maps wavefronts $\varphi$ to SH-WFS measurements
\begin{equation*}
s = \begin{pmatrix}s^x\\s^y\end{pmatrix}=\begin{pmatrix}\Gamma^x\varphi\\\Gamma^y\varphi\end{pmatrix} \eqqcolon \Gamma \varphi.
\end{equation*}  
Assuming that the wavefronts are described by piecewise continuous bilinear functions with nodal values $\varphi_{ij}$ we obtain the SH measurements in a subaperture $\Omega_{ij}$ by
\begin{equation*}
s_{ij}^x=\frac{(\varphi_{i,j+1}-\varphi_{i,j})+(\varphi_{i+1, j+1} - \varphi_{i+1,j})}{2},
\end{equation*}
\begin{equation*}
s_{ij}^y=\frac{(\varphi_{i+1,j}-\varphi_{i,j})+(\varphi_{i+1, j+1} - \varphi_{i,j})}{2}.
\end{equation*}

The wavefront aberrations $\varphi$ in the direction of a NGS are defined by 
\begin{equation*}
\varphi_\theta(x) = (P_\theta^{NGS}\phi)(x):=\sum_{\ell = 1}^L\phi_\ell(x+\theta h_\ell),
\end{equation*}
where $x=(x_1,x_2,0)$ is a point on the aperture, $\theta = (\theta_1,\theta_2,1)$ is the direction of the guide star and $h_\ell$ is the layer height. For a LGS at a fixed height $H$ the cone-effect has to be taken into account leading to
\begin{equation*}
\varphi_\theta(x) = (P_\theta^{LGS}\phi)(x):=\sum_{\ell = 1}^L\phi_\ell\left((1-\frac{h_\ell}{H})x+\theta h_\ell\right).
\end{equation*}
For details about the definition of the geometric propagation operator, either for NGS or LGS, we refer to \cite{Fusco}. 

The underlying mathematical problem is ill-posed \cite{Davison83}, i.e., there is an unstable relation between the measurements and the solution. In the AO-literature it is common to use the Bayesian framework for regularization, which allows the incorporation of statistical information regarding the turbulence model and noise. We define $\boldsymbol{S}$ and $\boldsymbol{\Phi}$ as random variables that correspond to the vectors of SH-WFS measurements and turbulent layers. Further, we assume the presence of noise modeled via a random variable $\boldsymbol{\eta}$. This leads to the following reformulation of Equation \eqref{eq:atmo} in the Bayesian framework 
\begin{equation}\label{eq:Bayesian}
\boldsymbol{S} = A\boldsymbol{\Phi} + \boldsymbol{\eta}.
\end{equation}

We assume the random variables to be Gaussian and use the maximum a posteriori (MAP) estimate to compute an optimal point estimate for the solution \cite{Fusco}. The solution of Equation \eqref{eq:Bayesian} is then given by
\begin{equation}\label{eq:MAP1}
x_{MAP} = \arg\min_{\phi \in \mathbb{R}^n} \{\Vert\phi\Vert^2_{C_\phi^{-1}} + \Vert s - A\phi\Vert^2_{C_\eta^{-1}}\},
\end{equation}
where $C_\phi^{-1}$ and $C_\eta^{-1}$ are the inverse covariance matrices of layers $\boldsymbol{\Phi}$ and noise $\boldsymbol{\eta}$, respectively. For details about the definition fo $C_\phi$ and $C_\eta$ we refer to \cite{vanKarman}. The norms in Equation \eqref{eq:MAP1} induced by the covariance matrices, which are symmetric and positive definite, are defined by
\begin{equation*}
\Vert x \Vert_C^2 := (Cx, x).
\end{equation*}
The solution to this minimization problem is given by the solution of the linear system of equations
\begin{equation}\label{eq:MAP}
(A^* C_\eta^{-1}A + C_\phi^{-1})\phi = A^* C_\eta^{-1}s.
\end{equation}

\section{Dual domain discretization}\label{sec:discretization}
In order to numerically compute a solution of Equation \eqref{eq:MAP} discretization is required. There are certain fundamental advantages of using wavelets, as already extensively studied for FEWHA in \cite{YuHeRa13b}. The main idea of the previously developed FEWHA is to use compactly supported orthonormal wavelets for representing the turbulent layers. The properties in the frequency domain allow a diagonal approximation of $C_\phi$. The atmospheric tomography operator $A$ has a more efficient representation in a finite element domain, where continuous piecewise bilinear functions are utilized to represent wavefronts and layers. 

\begin{figure}
	\centering
	\begin{tikzpicture}	
	\foreach \i in {1, ..., 5} {
		\draw[] (\i,1) -- (\i,5);
		\draw[] (1,\i) -- (5,\i);
	}
	\foreach \i in {1, ..., 5} {
		\foreach \j in {1, ..., 5} {
			\filldraw (\i,\j) circle [radius=0.1];
		}
	}
	
	\foreach \i in {1, ..., 5} {
		\draw[dashed, color=red] (\i*0.5+1.2,1.7) -- (\i*0.5+1.2,3.7);
		\draw[dashed, color=red] (1.7,\i*0.5+1.2) -- (3.7,\i*0.5+1.2);
	}
	\foreach \i in {1, ..., 5} {
		\foreach \j in {1, ..., 5} {
			\filldraw[color=red] (\i*0.5+1.16,\j*0.5+1.16) rectangle ++(2pt,2pt);
		}
	}
	\end{tikzpicture}
	\caption{Square grid of layers $\Omega_l$ in black with $2^{2J_l}=2^4$ points and equidistant spacing. Projected grid of subapertures $\Omega$ in red with $n_s^2=16$ subapertures.}
	\label{fig:grids}
\end{figure}
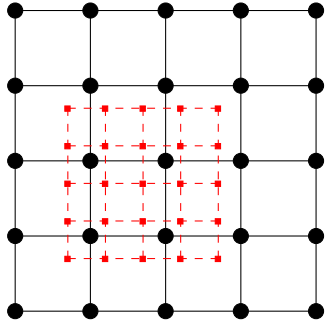

We use the same dual domain discretization approach for our method. We utilize a square grid with equidistant spacing on $\Omega$, the subaperture domain at the telescope pupil, to define the piecewise bilinear wavefront functions. The piecewise bilinear layer functions are defined using a square mesh with equidistant spacing on $\Omega_l$, the domain on which the turbulent layers are defined. This mesh consists of $2^{2J_l}$ points, where $J_l$ denotes the number of wavelet scales. See Figure \ref{fig:grids} for a graphical representation of $\Omega$ and $\Omega_l$. We obtain the following discretized MAP estimate
\begin{equation}\label{eq:fewha}
(\boldsymbol{W}^{-T}\hat{A}^TC_\eta^{-1}\hat{A}\boldsymbol{W}^{-1} + \alpha D)c=\boldsymbol{W}^{-T}\hat{A}^TC_\eta^{-1}s,
\end{equation}
where $\hat{A}$ is the atmospheric tomography operator in the finite element domain. We denote the linear mapping between the finite element and the wavelet domain by $\boldsymbol{W} = diag(\delta_1W,...,\delta_LW)$, where $W$ is the discrete wavelet transform and $\delta_l$ is the scaling constant at layer $l$. The operator $C_\eta^{-1}$ denotes the inverse covariance matrix of the noise and $D$ is a diagonal approximation of $C_\phi^{-1}$ in the frequency domain. We introduce a scalar factor $\alpha$ for tuning the balance between the fitting and the regularizing terms, see \cite{HeYu13} for more details. The vector $c$ is a concatenation of all wavelet coefficients of all turbulence layers and the vector $s$ is the concatenation of all SH sensor measurements from all guide star directions. In all our simulations we use periodic Daubechies-$3$ wavelets, which are an orthogonal wavelet family with compact support. 

For the sake of simplicity, we define the left-hand side operator of Equation \eqref{eq:fewha} by
\begin{equation}\label{eq:M}
\mathbf M := (\boldsymbol{W}^{-T}\boldsymbol{\hat{A}}^T\boldsymbol{\hat{C}}_\eta^{-1}\boldsymbol{\hat{A}}\boldsymbol{W}^{-1} + \alpha \mathbf D)
\end{equation}
and the right-hand side as
\begin{equation*}\label{eq:b}
b := \boldsymbol{W}^{-T}\boldsymbol{\hat{A}}^T \boldsymbol{C}_\eta^{-1}s.
\end{equation*}
Note, that the matrix $\mathbf M \in \mathbb{R}^{2^{2J_{\ell}}L\times 2^{2J_{\ell}}L}$ is symmetric and positive definite.

\section{Preconditioned conjugate gradient method for\\atmospheric tomography}\label{sec:pcg}
We are concerned with the case where the matrix $\mathbf M$ in Definition \eqref{eq:M} is very large, in particular, too large to make direct solution methods, such as the MVM, attractive or feasible. A direct solver requires the factorization of the left-hand side operator, which is computationally very expensive for large matrices. We focus on the most prominent iterative solver, for $\mathbf M$ being symmetric and positive definite, called conjugate gradient (CG) algorithm. An iterative method applies the forward operator $\mathbf M$ recursively to obtain the solution. Since all the operators involved in the definition of $\mathbf M$ in \eqref{eq:M} have a sparse structure and can be represented in a matrix-free way (see \cite{Yu14} for details), iterative methods are especially efficient here. The CG algorithm was already used for the previously developed FEWHA \cite{YuHeRa13b}. The most time consuming part, which is the application of $\mathbf M$, is well-parallelizable. For more details on the parallelization possibilities for FEWHA we refer to \cite{stadler2020}. 

The idea behind the CG method is that for a symmetric and positive definite matrix $\mathbf M$ solving the linear system $\mathbf Mc=b$ is equivalent to minimize $\frac{1}{2}(\mathbf Mc,c)-(b,c)$. This minimization is performed iteratively over a subspace spanned by the search directions $p$. Theoretically, the CG algorithm needs at most $2^{2J_\ell}L$ iterations, corresponding to the size of $\mathbf M$, until the exact solution is obtained. However, this would be far too much to meet the real-time requirements of the ELT. In practice, the method provides already a good approximation after a few iterations. Therefore, the number of CG iterations is often fixed for real-time applications, i.e., the algorithm terminates after a predetermined number of $maxIter$ iterations. Note, that under these circumstances the convergence of the method is not guaranteed. To improve convergence we use the residual vector $r_0=b-\mathbf Mc_0$, with an initial guess $c_0$, as a starting value of the CG method. In the community of AO this is referred to as warm restart. The convergence behavior of the CG method is affected by the eigenvalue distribution of the operator, which can be modified by preconditioning. We prefer a mixed preconditioner $\mathbf J^{-1} = (\mathbf J^{-1/2})^T \mathbf J^{-1/2}$ here to preserve the symmetry of the matrix, and thus be still able to apply the CG method. We obtain the following preconditioned version of Equation \eqref{eq:fewha}
\begin{equation}\label{eq:precond}
(\mathbf J^{-1/2})^T \mathbf M \mathbf J^{-1/2}y=(\mathbf J^{-1/2})^Tb.
\end{equation}
The solution $c$ to the original problem is given by
\begin{equation*}
c = \mathbf J^{-1/2}y.
\end{equation*}
The Jacobi preconditioner is a very simple way of preconditioning. It is a diagonal matrix with the diagonal components of $\mathbf J = diag(\mathbf M)$ and easy to invert. Within FEWHA a modified Jacobi preconditioner is utilized with a different weighting of the low and high frequency regimes \cite{YuHeRa13}. The benefit of such an approach is the reduction of CG iterations and an increased robustness and stability of the whole method. Moreover, this preconditioner is formulated as a diagonal matrix, which is very efficient in terms of computational costs. Algorithm~\ref{alg:PCG} shows the PCG method with the split Jacobi preconditioner $\mathbf J^{-1} = \mathbf J^{-1/2} \mathbf J^{-1/2}$. 

\begin{algorithm}[ht]
	\caption{PCG Algorithm}
	\small
	\label{alg:PCG}
	\begin{algorithmic}[1]
		\STATE{\textbf{Input:}\quad\quad~$c_0$ (previous wavelet coefficients)\\
			\quad\quad\quad\quad\quad~$r_0$ (previous residual vector)\\
			\quad\quad\quad\quad\quad~$maxIter$ (number of PCG iterations)\\
			\quad\quad\quad\quad\quad~$\mathbf J^{-1}$ (preconditioner)\\
		}
		\STATE{\textbf{Output:}\quad~$c^{(i+1)}$ (new wavelet coefficients)\\
			\quad\quad\quad\quad\quad~$r^{(i+1)}$ (new residual)}
		\STATE $p_0 = r_0$\\
		\STATE $z_0 = \mathbf J^{-1}r_0$
		\vspace{0.3cm}
		\FOR{$k=0,...,maxIter$}
		\STATE $q_k=\mathbf Mp_k$\\
		\STATE $\alpha = (r_k,z_k)/(p_k,q_k)$\\
		\STATE $c_{k+1} = c_k + \alpha p_k$, $r_{k+1} = r_{k}-\alpha q_k$\\
		\STATE $z_{k+1} = \mathbf J^{-1}r_{k+1}$\\
		\STATE $\beta=(r_{k+1},z_{k+1})/(r_{k},z_{k})$
		\STATE $p_{k+1} = z_{k+1}+\beta p_k$
		\ENDFOR
		\vspace{0.3cm}
		\STATE $c^{(i+1)} = c_{k+1}$, $r^{(i+1)} = r_{k+1}$
	\end{algorithmic}
\end{algorithm}

In the following we list some important properties of the CG method, see e.g.  \cite{AugCG}, which we will use in subsequent sections. The residuals are orthogonal to each other
\begin{equation*}
(r_i, r_j) = 0 \text{ for }i,j\geq 0, i\neq j.
\end{equation*}
The vectors $p_i$ are called search directions and are $M$-orthogonal to each other
\begin{equation*}
(\mathbf J^{-1/2}\mathbf M \mathbf J^{-1/2} p_i, p_j) = 0 \text{ for }i,j\geq 1, i\neq j.
\end{equation*}
Let us assume we have performed $m$ CG-iterations. We define $\mathbf R_m:=(r_0, ..., r_m)$ as the matrix of residuals and $\mathbf P_m := (p_0,...,p_m)$ as the matrix of conjugate search directions. The following conditions hold
\begin{equation}\label{eq:PD}
\mathbf P_m^T\mathbf M \mathbf P_m = \mathbf D_m \text{ and } Span(\mathbf R_m) = Span(\mathbf P_m)=\mathcal{K}_m(\mathbf J^{-1/2}\mathbf M\mathbf J^{-1/2},r_0),
\end{equation}
where $D_m$ is a diagonal matrix of size $m \times m$ and $\mathcal{K}_m(\mathbf J^{-1/2}\mathbf M\mathbf J^{-1/2},r_0)$ is the Krylov subspace of size $m+1$ generated by the initial residual $r_0$. Further, we introduce
\begin{equation}\label{eq:H}
\mathbf H_m = \mathbf I - \mathbf P_m \mathbf D_m^{-1}(\mathbf M \mathbf P_m)^T,
\end{equation}
the matrix of the M-orthogonal projection onto $\mathcal{K}_m(\mathbf J^{-1/2}\mathbf M\mathbf J^{-1/2},r_0)^\perp$.

\subsection{Krylov subspace recycling}\label{sec:recycling}
Within AO we are dealing with several right-hand sides $b$ of Equation \eqref{eq:fewha}, available consecutively in each time step. The CG method depends on the right hand side $b$ and has to be recomputed, in theory, whenever $b$ changes, i.e., in every single loop iteration. This is costly in terms of computational speed compared to direct solvers, where the factorization is independent from the right-hand side. However, within atmospheric tomography we can exploit the fact that the right hand side only changes slightly in every time step. The basic idea is to speed up the convergence of the current time step by using the search directions in $\mathbf P_m$ obtained from the previous system.

We define Equation \eqref{eq:precond} for several right-hand sides, which correspond to the WFS measurements from time steps $i=1,2,...$, by
\begin{equation}\label{eq:severalRHS}
\mathbf J^{-1/2}\mathbf M\mathbf J^{-1/2}c^{(i)}=\mathbf J^{-1/2}b^{(i)},
\end{equation}
where $b^{(i)}:= \boldsymbol{W}^{-T}\boldsymbol{\hat{A}}^T \boldsymbol{C}_\eta^{-1}s^{(i)}$. The main idea is to use the information obtained when solving Equation \eqref{eq:severalRHS} for a certain time step $i$ with $m$ iterations of the PCG method for the upcoming time steps and recycle the Krylov subspace $\mathcal{K}_m(\mathbf J^{-1/2}\mathbf M\mathbf J^{-1/2},r_0^{(i)})$ generated in the previous system.

A first idea in that direction is to improve the convergence of the PCG method by choosing a more optimized initial guess $c_0$ as a starting value. In fact, $c_0^{(i+1)}$ can be chosen such that the initial residual $r_0^{(i+1)}$ is orthogonal to the Krylov subspace $\mathcal{K}_m(\mathbf J^{-1/2}\mathbf M \mathbf J^{-1/2},r_0^{(i)})$ generated in the previous time step. This Galerkin projection technique was first proposed by Saad in \cite{Saad87} for the Lanczos process and adapted in \cite{AugCG} for the CG method. We enforce this orthogonality condition by
\begin{equation*}
\mathbf P_m^Tr_0^{(i+1)} = 0.
\end{equation*}
To satisfy this condition we choose
\begin{equation*}
c_0^{(i+1)} = c_0^{(i)} + \mathbf P_m \mathbf D_m^{-1}\mathbf P_m^Tr_0^{(i+1)},
\end{equation*}
where $\mathbf P_m$ and $\mathbf D_m$ are defined by Equation \eqref{eq:PD}. For the proof we refer to \cite{AugCG}.

Another way to improve the convergence behavior is to keep this orthogonality throughout the iterations of the PCG method. As before, this method was first proposed for the Lanczos process in \cite{Saad87} (called modified Lanczos process) and adapted for the CG method in \cite{AugCG} (referred to as augmented CG). Within this approach we use two subspaces: the Krylov subspace generated for the first system $\mathcal{K}_m(\mathbf J^{-1/2}\mathbf M \mathbf J^{-1/2},r_0^{(i)})$ and the subspace $Span(r_0^{(i+1)},...,r_k^{(i+1)})$. The residual $r_{k+1}^{(i+1)}$ must be orthogonal to both subspaces and the directions $p_{k+1}$ must be $\mathbf M$-orthogonal two both subspaces. Hence, this method is a projection method onto the space
\begin{equation*}\label{eq:Kmk}
\mathcal{K}_{m,k}(\mathbf J^{-1/2} \mathbf M \mathbf J^{-1/2},r_0^{(i)},r_0^{(i+1)}) = \mathcal{K}_m(\mathbf J^{-1/2} \mathbf M \mathbf J^{-1/2},r_0^{(i)}) + Span(r_0^{(i+1)},...,r_k^{(i+1)}),
\end{equation*}
which is not a Krylov-subspace. 

The projection is defined by the following three conditions , see \cite{AugCG}:
\begin{enumerate}
	\item $p_0^{(i+1)} = (\mathbf I-\mathbf P_m \mathbf D^{-1}(\mathbf M \mathbf P_m)^T)r_0^{(i+1)}$
	\item $c_{k+1}-c_k\in \mathcal{K}_{m,k}(\mathbf J^{-1/2} \mathbf M \mathbf J^{-1/2},r_0^{(i)},r_0^{(i+1)})$
	\item $(r^{(i+1)}_{k+1},z) = 0 \text{ for all } z\in \mathcal{K}_{m,k}(\mathbf J^{-1/2} \mathbf M \mathbf J^{-1/2},r_0^{(i)},r_0^{(i+1)})$
\end{enumerate}
The last condition is called Petrov-Galerkin condition and is satisfied if the current residual $r^{(i+1)}_{k+1}$ is orthogonal to $r^{(i+1)}_k$ and the current direction $p^{(i+1)}_{k+1}$ is $\mathbf M$-conjugate to $p^{(i+1)}_{k}$ and $p^{(i)}_{m}$. 

Algorithm \ref{alg:augPCG} shows the augmented PCG method. The method starts with computing an initial guess $c^{(i+1)}_{0}$ (Line \ref{line:init}) such that the initial residual $r^{(i+1)}_0$ is orthogonal to the Krylov subspace $\mathcal{K}_m(\mathbf J^{-1/2} \mathbf M \mathbf J^{-1/2},r_0^{(i)})$. Further, the initial descent direction $p^{(i+1)}_{0}$ is calculated in lines \ref{line:initDir1} - \ref{line:initDir2} such that it is conjugate to the descent directions in $\mathbf P_m$. In each PCG iteration the new search direction $p_k^{(i+1)}$ is orthogonalized against the last search direction of the previous system $p_m^{(i)}$ in Line \ref{line:newSearchDir}. 

\begin{algorithm}[ht]
	\caption{Augmented PCG Algorithm \cite{AugCG}}
	\small
	\label{alg:augPCG}
	\begin{algorithmic}[1]
		\STATE{\textbf{Input:}\quad\quad~$c_0$, $r_0$ (previous wavelet coefficients and residual)\\
			\quad\quad\quad\quad\quad~$maxIter$ (number of PCG iterations)\\
			\quad\quad\quad\quad\quad~$\mathbf J^{-1}$ (preconditioner)\\
			\quad\quad\quad\quad\quad~$p^{(i)}$, $q^{(i)}$ (previous descent directions)
		}
		\STATE{\textbf{Output:}\quad~$c^{(i+1)}$, $r^{(i+1)}$ (new wavelet coefficients and residual)\\
			\quad\quad\quad\quad\quad~$p^{(i+1)}$, $q^{(i+1)}$ (new descent directions)}
		\FOR{$j=0,...,m$}
		\STATE $\sigma_j = (r,p_j^{(i)})/(p_j^{(i)},q_j^{(i)})$\\
		\vspace{0.1cm}
		\STATE $c_{0}=c_{0}+\sigma_j p_j^{(i)}$, $r_{0} = r_{0}-\sigma_j q_j^{(i)}$ \label{line:init}
		\ENDFOR
		\vspace{0.3cm}
		
		\STATE $z_0 = \mathbf J^{-1}r_{0}$\\\label{line:initDir1}
		\FOR{$j=0,...,m$}
		\STATE $z_0 = z_0 - (z,q_j^{(i)})/(p_j^{(i)},q_j^{(i)})$
		\ENDFOR
		\vspace{0.3cm}
		\STATE $p_0^{(i+1)} = z_0$\\\label{line:initDir2}
		\FOR{$k=0,...,maxIter$}
		\STATE $q_k^{(i+1)}=\mathbf Mp_k^{(i+1)}$\\
		\vspace{0.1cm}
		\STATE $\alpha = (r_{k},z_k)/(p_k^{(i+1)},q_k^{(i+1)})$\\
		\vspace{0.1cm}
		\STATE $c_{k+1} = c_{k} + \alpha p_k^{(i+1)}$, $r_{k+1} = r_{k}-\alpha q_k^{(i+1)}$\\
		\vspace{0.1cm}
		\STATE $z_{k+1} = \mathbf J^{-1}r^{(i+1)}_{k+1}$\\
		\vspace{0.1cm}
		\STATE $\mu = (z_{k+1},q_m^{(i)})/(p_m^{(i)},p_m^{(i)})$\\
		\vspace{0.1cm}
		\STATE $z_{k+1} = z_{k+1}-\mu p_m^{(i)}$\\
		\vspace{0.1cm}
		\STATE $\beta=(r_{k+1},z_{k+1})/(r_{k},z_{k})$
		\vspace{0.1cm}
		\STATE $p_{k+1}^{(i+1)} = z_{k+1}+\beta p_k^{(i+1)}$\label{line:newSearchDir}
		\ENDFOR
		\vspace{0.3cm}
		
		\STATE $c^{(i+1)}=c_{k+1}$, $r^{(i+1)}=r_{k+1}$
	\end{algorithmic}
\end{algorithm}

\section{Augmented Finite Element Wavelet Hybrid Algorithm}\label{sec:augFEWHA}
Our augmented FEWHA differs from the original algorithm proposed in \cite{Yu14} only in the fact that we use the augmented PCG algorithm described in Algorithm \ref{alg:augPCG} instead of the classical PCG method. Algorithm \ref{alg:augFEWHA} shows the augmented wavelet reconstructor for time step $(i+1)$. The main input is the measurement vector $s^{(i+1)}$ and the output is the new shape of the mirrors $a^{(i+1)}$. The AO system can either operate in closed or in open loop. If open loop control is applied, the measurements are directly obtained from the wavefronts, whereas in closed loop control the pseudo open loop measurements are calculated as a first step of the algorithm in Line \ref{line:openLoop}. Due to the two-step delay (see \cite{Poettinger_2019} for details), the DM shape from the previous step is used. The right-hand side $b^{(i+1)}$ is computed in Line \ref{line:RHS} with the new measurement vector $s^{(i+1)}$ and, subsequently, the initial residual $r_0^{(i+1)}$ is updated in Line \ref{line:res}. The atmospheric reconstruction takes place in Line \ref{line:PCG}, where $p^{(i)}$ and $q^{(i)}$ are used within the PCG method to improve the solution by projection. Afterwards, the layers are fitted to actuator commands in Line \ref{line:fitting} and the closed or open loop control is applied in lines \ref{line:closedcontrol} - \ref{line:opencontrol}. The new DM shapes are a linear combination of the current and the reconstructed DM shapes, weighted by a scalar value called gain, between zero and one. This gain control improves the stability of the reconstruction. For closed loop control the artificially added DM shapes $a^{(i-1)}$ are subtracted from the computed mirror shapes $\tilde{a}$, such that the difference is simply the reconstruction from the closed loop measurements.

\begin{algorithm}[ht]
	\caption{Augmented Wavelet Reconstructor}
	\small
	\label{alg:augFEWHA}
	\begin{algorithmic}[1]
		\STATE{\textbf{Input:}\quad\quad~$s^{(i+1)}=(s_g)^G_{g=1}$ (measurement vector)\\
			\quad\quad\quad\quad\quad~$gain$ (scalar weight)\\
			\quad\quad\quad\quad\quad~$c^{(i)}$ (previous wavelet coefficients)\\
			\quad\quad\quad\quad\quad~$b^{(i)}$ (previous right-hand side)\\
			\quad\quad\quad\quad\quad~$r^{(i)}$ (previous residual vector)\\
			\quad\quad\quad\quad\quad~$a^{(i-1)}, a^{(i)}$ (previous two DM shape)\\
			\quad\quad\quad\quad\quad~$maxIter$ (number of PCG iterations)\\
			\quad\quad\quad\quad\quad~$J^{-1/2}$ (preconditioner)\\
			\quad\quad\quad\quad\quad~$p^{(i)}$,$q^{(i)}$ (previous descent directions)
		}
		\STATE{\textbf{Output:}\quad~$a^{(i+1)}$ (next DM shape)}
		
		\IF{loop = closed} \label{line:openLoop}
		\STATE $s^{(i+1)} = s^{(i+1)} + \mathbf \Gamma a^{(i-1)}$\label{alg.LTAO_MOAO.applyH}
		\ENDIF
		\vspace{0.3cm}
		
		\STATE $b^{(i+1)} = \mathbf W^{-T} {\mathbf A}^T \mathbf C_{\eta}^{-1} s^{(i+1)}$ \label{line:RHS}
		\STATE $r_0 = b^{(i+1)} - \mathbf M c^{(i)} = (b^{(i+1)} - b^{(i)}) + r^{(i)}$ \label{line:res} 	
		\vspace{0.3cm}				
		\STATE $[c^{(i+1)}, r^{(i+1)}, p^{(i+1)}, q^{(i+1)}] = augPCG(c^{(i)}, r_0, \mathbf J^{-1/2}, p^{(i)}, q^{(i)}, maxIter)$\\\label{line:PCG}
		\vspace{0.3cm}
		\STATE$\tilde a = \mathbf F \mathbf W^{-1} c^{(i+1)}$\label{line:fitting}
		
		\vspace{0.3cm}
		\IF{loop = closed}\label{line:closedcontrol}
		\STATE $a^{(i+1)} = a^{(i)} + \mathrm{gain} \cdot (\tilde a - a^{(i-1)})$ 
		\ELSIF{loop = open}
		\STATE $a^{(i+1)} = (1 - \mathrm{gain}) \cdot a^{(i)} + \mathrm{gain} \cdot \tilde a$ 
		\ENDIF \label{line:opencontrol}
	\end{algorithmic}
\end{algorithm}

\section{Quality evaluation}\label{sec:numerics}
We evaluate the quality of our proposed method using the Strehl ratio, which is a frequently used quality evaluation criterion within AO. The short exposure (SE) Strehl is defined as the ratio between the real energy distribution of incoming light in the image plane $I(x,y)$ over the hypothetical distribution $I_D(x,y)$, which stems from the assumption of diffraction-limited imaging,
\begin{equation*}
S := \frac{max_{(x,y)}I(x,y)}{max_{(x,y)}I_D(x,y)} \in[0,1].
\end{equation*}
The higher the Strehl ratio the better the quality of the AO system. The maximum of $1$ is reached only in the diffraction-limited case. A tip/tilt-mode, which leads to a horizontal shift of $I(x,y)$, does not influence $S$. In order to detect a tip/tilt mode, the long exposure (LE) Strehl ratio is used. The LE Strehl represents the quality of the image from the start of the loop until a certain time step. We use the LE Strehl in the K band, i.e., for a wavelength of $2200$~nm. The short and long exposure Strehl ratio is often indicated in $\%$. For more details about the Strehl ratio we refer to \cite{Roddier}.

\subsection{ELT simulations}
\begin{figure}[t]
	\centering
	\footnotesize
	\renewcommand{\arraystretch}{1.3}
	\begin{tabular}[b]{|l|l|}
		\hline
		\textbf{Parameter} & \textbf{Value} \\\hline\hline
		Telescope diameter $D$ &  $39$ m\\
		Central obstruction & $\sim 28~\%$\\
		Fried parameter $r_0$ & $0.129$ m\\
		Na-layer thickness $FWHM$ & $11.4$ km\\
		Na-layer height $H$ & $90$ km\\
		Evaluation criterion & LE Strehl\\
		Evaluation wavelength & K band\\
		Gain & 0.4\\
		Number of wavelet scales $J_\ell$ & $6$\\
		Regularization parameter $\alpha$ & 2 \\
		Number of loop iterations & $500$\\\hline
	\end{tabular}
	\qquad
	\begin{tikzpicture}[scale=0.6]
	\draw [very thin, lightgray] (0,0) grid (6,6);
	\draw (-0.5,1.2) node[below] {$-60$};
	\draw (-0.5,2.2) node[below] {$-30$};
	\draw (-0.5,3.2) node[below] {$0$};
	\draw (-0.5,4.2) node[below] {$30$};
	\draw (-0.5,5.2) node[below] {$60$};
	\draw (0.9, 0) node[below] {$-60$};
	\draw (1.9, 0) node[below] {$-30$};
	\draw (2.9, 0) node[below] {$0$};
	\draw (3.9,0) node[below] {$30$};
	\draw (4.9,0) node[below] {$60$};
	\foreach \x in {1,2,3,4,5}
	\draw [red] plot [only marks, mark size=1.5, mark=*] coordinates {(\x,1) (\x,2) (\x,3) (\x,4) (\x,5)};
	\draw [blue] plot [only marks, mark size=5, mark=diamond] coordinates {(2,4.666) (4,4.666) (1, 3) (5, 3) (2,1.333) (4,1.333)};
	\draw [green] plot [only marks, mark size=3.5, mark=square] coordinates {(0.55,3) (4.2,0.8) (4.2,5.2)};
	
	\begin{customlegend}[
	legend entries={
		quality evaluation grid,
		natural guide star, 
		laser guide star,
	},
	legend style={at={(6,9)},font=\footnotesize}]
	\addlegendimage{only marks, mark=*, color=red, mark size=1.5}
	\addlegendimage{only marks, mark=square, color=green, mark size=2}
	\addlegendimage{only marks, mark=diamond, color=blue, mark size=3}
	\end{customlegend}
	\end{tikzpicture}
	\caption{General parameter setting for ELT test case and the star asterism of MCAO NGS and LGS setting with the $5 \times 5$ quality evaluation grid.}
	\label{fig:gerneralSettings}
\end{figure}
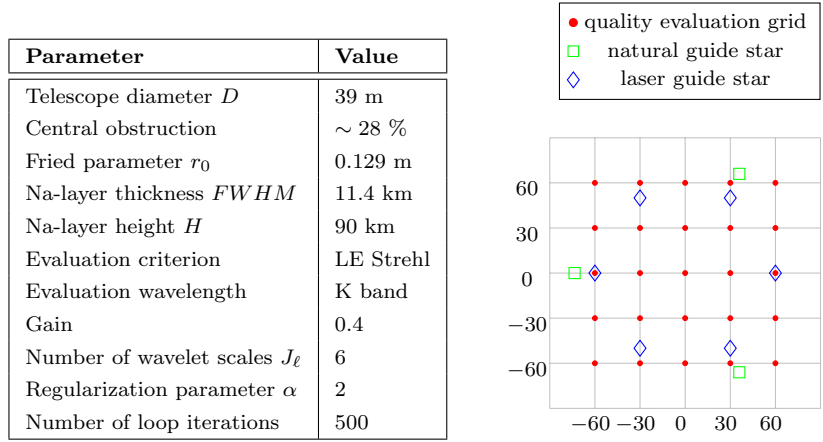

To verify the quality and computational performance of the proposed algorithm we use the design of the ELT with an assumed telescope diameter of $39$~m, where about $28~\%$ are obstructed. Our numerical simulations are executed via the software package MOST, see e.g. \cite{Au17}, which was developed in-house as an alternative to OCTOPUS \cite{LVK06}. We model the vertical density profile of the laser beam scatter in the sodium layer by a Gaussian random variable with mean $H=90$~km and a full width at half maximum (FWHM) of $11.4$~km. The table on the left side of Figure \ref{fig:gerneralSettings} summarizes the general parameter settings for our numerical simulations. We examine the performance of FEWHA versus its augmented version for the AO systems LTAO, MOAO and MCAO. All these systems use atmospheric tomography to correct the optical distortions. For the LTAO and MOAO configuration we assume to have $3$ WFS with $80 \times 80$ subapertures assigned to $3$ NGS positioned in a circle of $10$ arcmin diameter. Further, we assume to have $6$ WFS with $80 \times 80$ subapertures assigned to $6$ LGS positioned in a circle of $7.5$ arcmin diameter. For the MCAO test case we have two WFS with $1 \times 1$ subapertures and one with $2 \times 2$ assigned to $3$ NGS positioned in a circle of $2$ arcmin diameter. The $6$ WFS with $80 \times 80$ subapertures are assigned to $6$ LGS positioned in a circle of $8/3$ arcmin diameter. A graphical representation of the MCAO star asterism is shown on the right side of Figure \ref{fig:gerneralSettings}. In the LTAO case we simulate a $9$ layer atmosphere and use a single DM with $81 \times 81$ actuators to correct for atmospheric distortions. The MOAO simulations have a similar setup, only the NGS and LGS flux differs from the LTAO setting. For the MCAO tests we simulate a $3$ and a $9$ layer atmosphere. In contrast to LTAO and MOAO, the MCAO system uses $3$ DMs with $81 \times 81$, $47 \times 47$ and $53 \times 53$ at different altitudes and actuator spacing. In the $3$ layer approach we reconstruct the atmosphere directly at the deformable mirrors, whereas in the $9$ layer case we solve the fitting equation using an unpreconditioned CG algorithm with 4 iterations. The same setup was already used in \cite{Yu14} for FEWHA. Table \ref{tab:SystemSettings} summarizes the system specific configurations.

\begin{table}[ht]
	\centering
	\footnotesize
	\renewcommand{\arraystretch}{1.3}
	\begin{tabular}{|l|l|l|l|}
		\hline
		\textbf{Parameter} & \textbf{LTAO} & \textbf{MOAO} & \textbf{MCAO} \\\hline\hline
		Number of Layer $L$ & $9$ & $9$ & $3$ or $9$\\
		Number of NGS $G_{NGS}$ & $3$ & $3$ & $3$\\
		\quad Number of WFS subapertures $n_{lay}$ & $84 \times 84$ & $84 \times 84$ & one $2 \times 2$, two $1 \times 1$\\
		\quad Direction, circle of & $10$ & $10$ & $8/3$ arcmin diameter\\
		\quad Wavelength & $589$~nm & $589$~nm & $589$~nm \\
		\quad Flux & $500$ & $5-500$ & $300$ photons/subap./frame\\
		Number of LGS $G_{LGS}$  & $6$ & $6$ & $6$\\
		\quad Number of WFS subapertures $n_{lay}$ &  $84 \times 84$ & $84 \times 84$ & $84 \times 84$\\
		\quad Direction, circle of & $7.5$ & $7.5$ & $2$ arcmin diameter \\
		\quad Wavelength & $500$~nm & $500$~nm & $1650$~nm\\
		\quad Flux & $50-500$ & $5-500$ & $50-500$ photons/subap./frame\\
		Number of DMs & $1$ & $1$ & $3$ \\
		\quad Number of actuators &  $85 \times 85$ & $85 \times 85$ & $85 \times 85$, $47 \times 47$, $53 \times 53$\\
		\quad DM altitude & $0$~m & $0$~m & $0$~m, $4$~km, $12.7$~m\\
		\quad Actuator spacing & $0.5$~m &  $0.5$~m  & $0.5$~m, $1$~m, $1$~m\\\hline
	\end{tabular}
	\caption{LTAO, MOAO and MCAO system specific parameters.}
	\label{tab:SystemSettings}
\end{table}

\begin{figure}[ht]
	\centering
	\captionsetup{justification=centering}
	\includegraphics[width=0.8\textwidth]{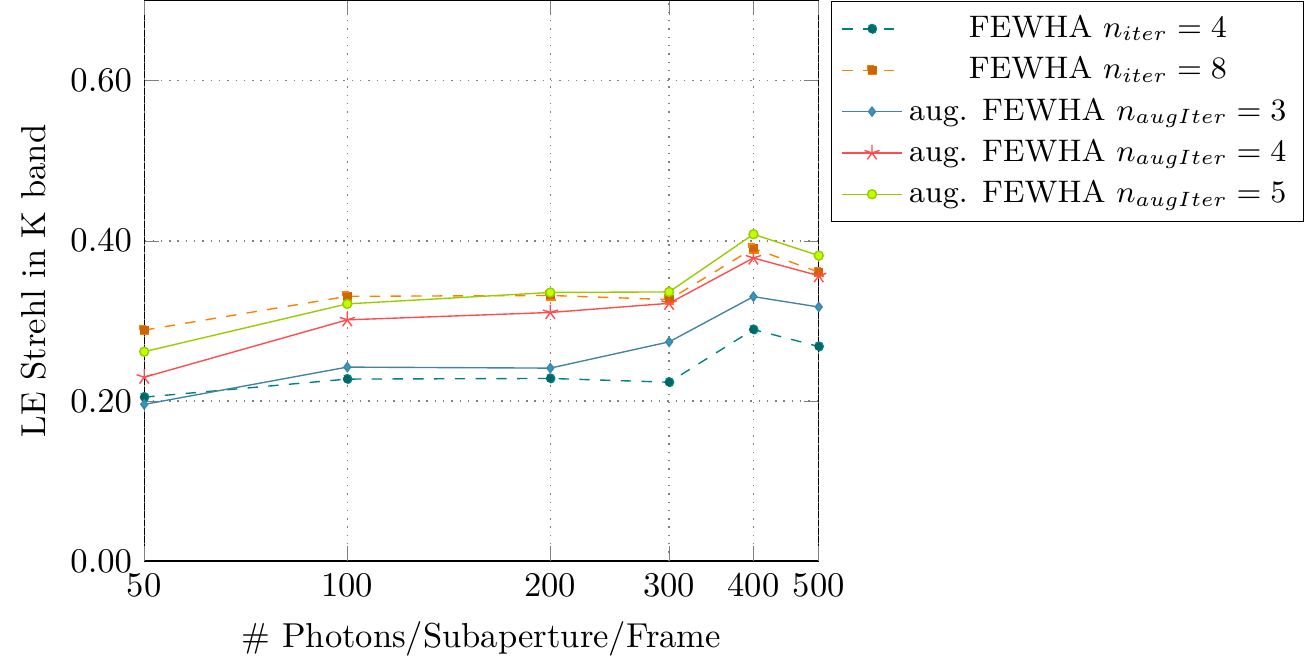}
	\caption{LTAO LE Strehl over the course of 500 iterations with different number of $n_{iter}$ and $n_{augIter}$ for FEWHA and its augmented version.}
	\label{fig:LTAO_LE}
\end{figure}

\begin{figure}[ht]
	\centering
	\captionsetup{justification=centering}
	\includegraphics[width=0.8\textwidth]{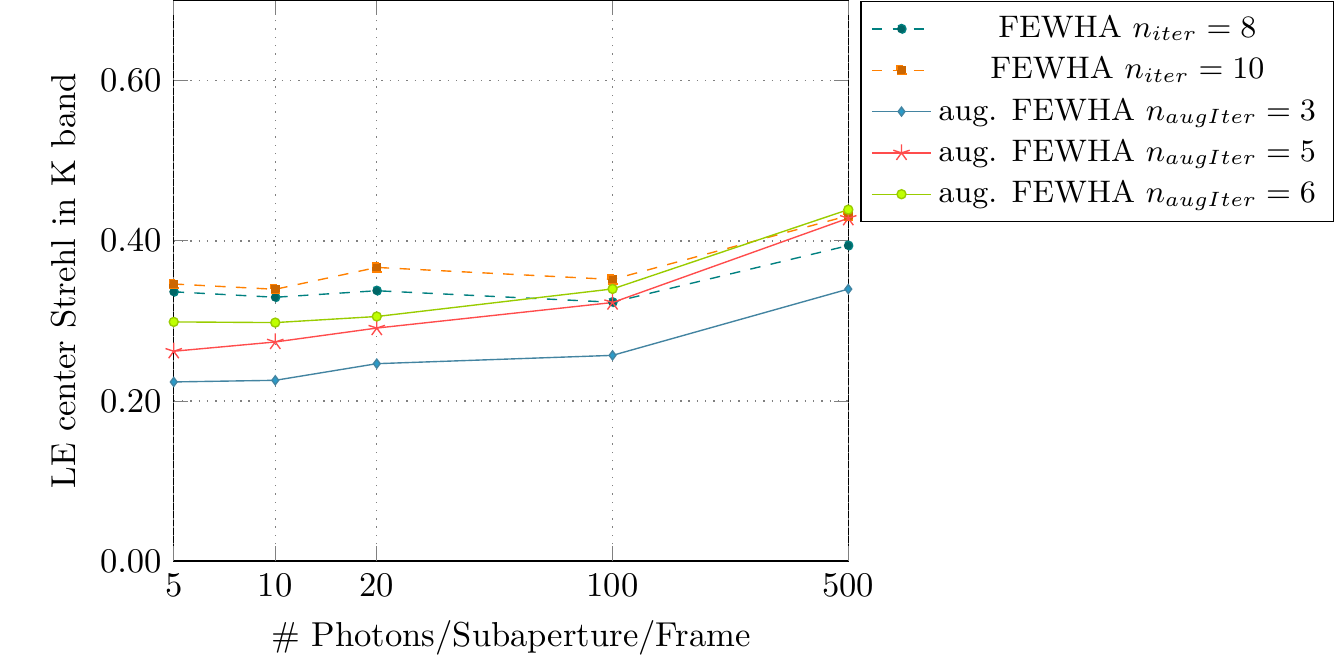}
	\caption{MOAO center LE Strehl over the course of 500 iterations with different number of $n_{iter}$ and $n_{augIter}$ for FEWHA and its augmented version.}
	\label{fig:MOAO_LE}
\end{figure}

\begin{figure}[ht]
	\centering
	\captionsetup{justification=centering}
	\includegraphics[width=1.0\textwidth]{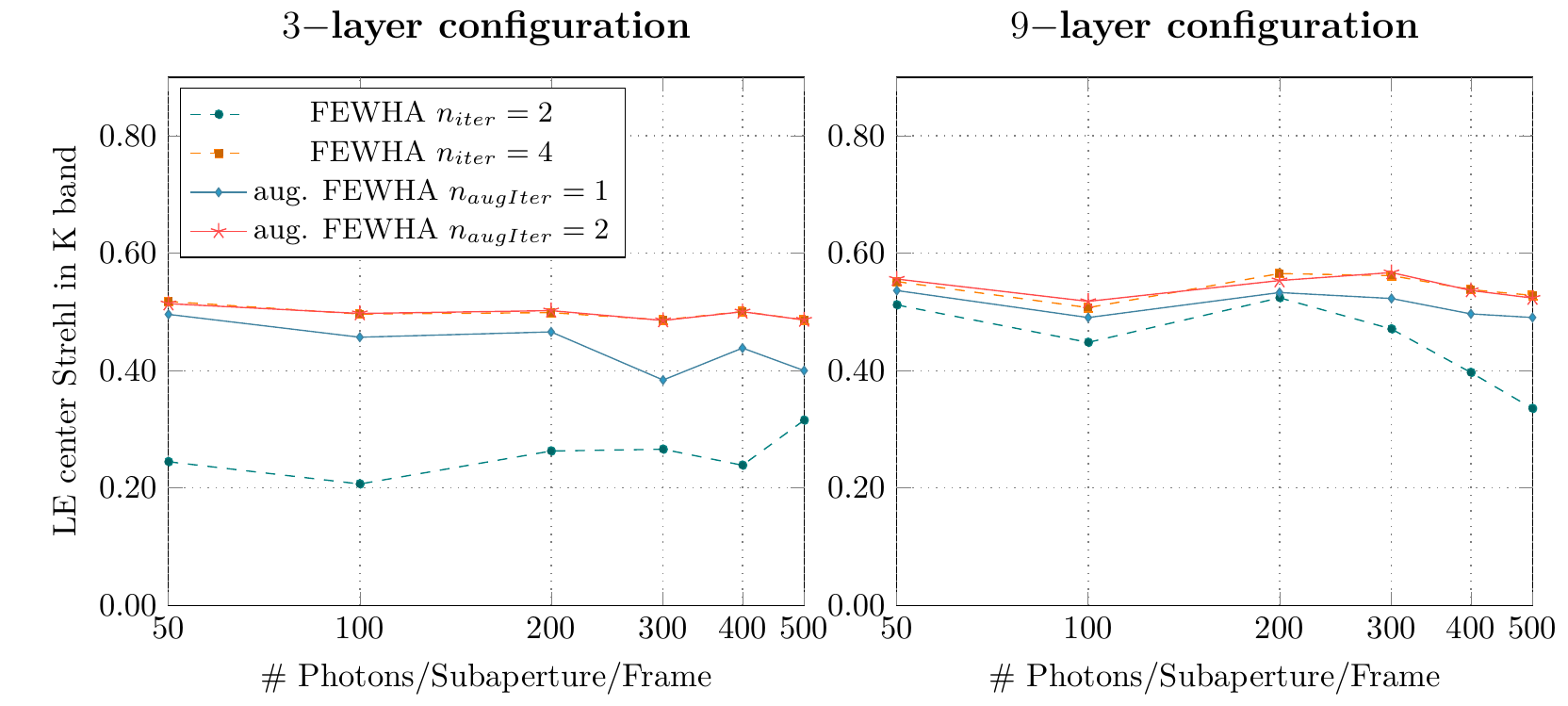}
	\caption{MCAO center LE Strehl over the course of 500 iterations for the $3$ layer configuration (on the left) and the $9$-layer setting (on the right) with different number of $n_{iter}$ and $n_{augIter}$ for FEWHA and its augmented version.}
	\label{fig:MCAO_LE}
\end{figure}

Figure \ref{fig:LTAO_LE} shows the LE Strehl for the LTAO simulations. For FEWHA we vary the number of PCG iterations $n_{iter}$ between $4$ and $8$, whereas for augmented FEWHA the number of iterations $n_{augIter}$ varies between $3$ and $5$. We observe that for a higher flux the classical FEWHA requires almost double the iterations compared to the augmented version to obtain a similar LE Strehl. For the low flux tests the situation is slightly different, however, also here we can see a reduction in the number of iterations. This leads to a non-negligible reduction of FLOPs for the augmented FEWHA (see Section \ref{sec:performance} for a detailed analysis), hence, a considerable speed-up of the whole algorithm. Figure~\ref{fig:MOAO_LE} presents the center LE Strehl for the MOAO simulations. We increase the number of FEWHA PCG iterations up to $n_{iter}=10$. This makes the augmented FEWHA even more efficient here, because saving approximately half the iterations saves more FLOPs as for the LTAO test case. Figure \ref{fig:MCAO_LE} shows the LE center Strehl for the MCAO $3$-layer setting on the left and the $9$-layer configuration on the right. Here, $n_{iter}=4$ is enough to meet the quality requirements. For augmented FEWHA we vary the number of iterations $n_{augIter}$ between $1$ and $2$. The augmented version with half the iterations provides the same quality as FEWHA also for the low flux cases. 

\subsection{Error analysis}\label{sec:error}
We are now interested in calculating an asymptotic error bound for the classical CG, the classical PCG and the augmented PCG at a certain iteration $k$. In \cite{AugCG} they show that the augmented CG computes the minimum over the subspace $\mathcal{K}_{m,k}(M,r_0^{(i)}, r_0^{(i+1)})$ and that this minimum is less or equal to the minimum over the subspace $\mathcal{K}_{k}(H^T M H, r_0^{(i+1)})$, where $H$ is defined as in Equation \eqref{eq:H}. Hence, the augmented CG is just a classical CG applied to the special matrix $H^T M H$ and the theory of convergence rates from the classical method can be applied. The augmented PCG can be seen as CG preconditioned by $HH^T$ and $J^{-1}$. Let $\kappa$ be the condition number of $H^TMH$, then we can use the theory of the classical CG, see e.g. \cite{Sluis1986}, to obtain the error bound for the augmented CG at iteration $k$ by
\begin{equation*}
\norm{c_k - c}_{M} \leq 2 \norm{c_0 - c}_{M} (\frac{\sqrt{\kappa}-1}{\sqrt{\kappa}+1})^k.
\end{equation*}
Note, that the the condition number of $H^TMH$ is less or equal to the condition number of $M$ \cite{AugCG}. Thus, the asymptotical convergence rate of the augmented PCG is less or equal to the asymptotic convergence rate of the classical PCG. We want to verify this statement for the MCAO test case defined in Table \ref{tab:SystemSettings}. We restrict ourself to this special test case, because it is related to MAORY, a first-light instrument of the ELT. The atmospheric tomography problem for MAORY is considerably complex, thus, it provides a suitable real-world example. For that purpose we calculate the condition numbers of $M$, $J^{-1/2}MJ^{-1/2}$ and $J^{-1/2}H^TMHJ^{-1/2}$ as
\begin{itemize}
	\item $\kappa_0:= cond(M) \approx 3\cdot 10^{8}$,
	\item $\kappa_1:=cond(J^{-1/2}MJ^{-1/2})\approx 2,76 \cdot 10^{7}$ and
	\item $\kappa_2:=cond(J^{-1/2}H^TMHJ^{-1/2}) \approx 1 \cdot 10^7$.
\end{itemize}
We can observe that $\kappa_0 > \kappa_1 > \kappa_2$. This might give an indication why we can decrease the number of PCG iterations for the augmented FEWHA compared to the original algorithm and still obtain a similar quality. See Figure \ref{fig:MCAO_LE} for a graphical representation of this behavior. 

\section{Computational performance}\label{sec:performance}
We compare FEWHA and its augmented version regarding floating point operations, memory usage and runtime on CPU. We use the LTAO, MOAO and MCAO 3-layer setting as defined in the previous section. We omit the 9-layer MCAO configuration here, because if we reconstruct more layers than DMs we have to solve an additional minimization problem, which is costly in terms of speed. Further, we fix $n_{iter}$ and $n_{augIter}$ as shown in Table \ref{tab:FEWHAELT_iter}, based on our observations from the numerical simulations in Section \ref{sec:numerics}, to obtain a similar quality. For details about FLOPs and memory usage for the MVM and FEWHA we refer to \cite{StaAO4ELT}. 

\begin{table}[ht]
	\center
	\footnotesize
	\renewcommand{\arraystretch}{1.3}
	\begin{tabular}{|c|c|c|}
		\hline
		\textbf{Test setting} & $n_{iter}$ & $n_{augIter}$ \\\hline
		\textbf{LTAO} & $8$ & $4$\\\hline
		\textbf{MOAO} & $10$ & $5$\\\hline
		\textbf{MCAO 3 layer} & $4$ & $2$\\\hline
	\end{tabular}
	\caption{Number of PCG iterations for FEWHA and augmented FEWHA.}
	\label{tab:FEWHAELT_iter}
\end{table}

\subsection{Floating point operations}
The FLOPs for FEWHA and its augmented version differ only in the PCG step. There are two additional parts for the augmented PCG: the projection of the initial guess $c_0$ and the initial residual $r_0$ onto the Krylov subspace of the previous system and the projection of the descent directions $p_k$ onto the last vector of the descent directions of the previous system $p_m$ in every time step. In every iteration one dot-product and one vector update is added. We want to stress that the augmented FEWHA does not introduce new matrix-vector products, which are the most time consuming steps of the algorithm. In the previous section we show that the asymptotic convergence rate of the augmented PCG is not higher than for the classical PCG. In fact, our numerical simulations in Section \ref{sec:numerics} showed that we can save approximately half the PCG iterations $n_{augIter} \approx n_{iter}/2$ with using the Krylov recycling technique. Hence, the overhead of the augmented FEWHA should be easily compensated by the reduction in the number of iterations. Table \ref{tab:FEWHAFLOPs} shows the theoretical FLOPs for augmented FEWHA. The first row corresponds to the classical PCG method within FEWHA, whereas the second and third line refer to the additional computations that have to be done within augmented PCG. In Table \ref{tab:FEWHAFLOPsELT} the results for the ELT configurations are presented. We can see immediately that we save approximately half the FLOPs when using the augmented PCG instead of the classical PCG. This is what we expect when choosing $n_{augIter}=n_{iter}/2$.  

\begin{table}[ht]
	\center
	\small
	\renewcommand{\arraystretch}{1.3}
	\begin{tabular}{|p{4cm}|p{6cm}|}
		\hline
		\textbf{Computation step} & \textbf{Theoretical FLOPs}\\\hline
		PCG iterations & ${\scriptstyle\{[14G_{LGS}+2G_{NGS}+12G]n_s^2+4G n_s}$\\ & ${\scriptstyle+[(L-1)(15.6G_{LGS}+18G_{NGS})+L(G-1)](n_s+1)^2}$\\ & ${\scriptstyle+L\frac{176}{3}(4n_{lay}^2-1)+13n_{lay}^2L\}n_{iter}}$\\\hline
		Initialization  &  ${\scriptstyle (9n_{lay}^2L)n_{augIter}}$\\
		Projection in every iteration & ${\scriptstyle (4n_{lay}^2L)n_{augIter}}$\\\hline
	\end{tabular}
	\caption{Theoretical FLOPs for FEWHA and augmented FEWHA.}
	\label{tab:FEWHAFLOPs}
\end{table}

\subsection{Memory usage}
For augmented FEWHA we have to save the descent directions $p_k$ and $q_k=Mp_k$ for $k=1,...,n_{augIter}$. Additionally, to decrease the number of FLOPs and avoid unnecessary re-computations we save the inner products $(p_k,q_k)$. Both vectors are of size $n_{lay}^2L$, hence, in total we need $(2n_{lay}^2L+1)n_{augIter}$ units of additional storage. The last column of Table \ref{tab:FEWHAFLOPsELT} shows the additional memory usage of augmented FEWHA over the classical version, assuming single precision (32 bit) floating point numbers. Compared to the overall memory usage of FEWHA and especially compared to the memory intensive MVM the additional memory usage for augmented FEWHA is almost negligible.

\begin{table}[ht]
	\center
	\footnotesize
	\renewcommand{\arraystretch}{1.3}
	\begin{tabular}{|p{2.3cm}|p{2.3cm}|p{2.3cm}|p{3.3cm}|}
		\hline
		\textbf{Test setting} & \textbf{FLOPs~PCG FEWHA} & \textbf{FLOPs PCG aug.~FEWHA} & \textbf{Add. memory usage aug.~FEWHA}\\\hline\hline
		\textbf{LTAO} & $376$~MFLOPs &  $196$~MFLOPs & $4.7$~MB\\\hline
		\textbf{MOAO} & $470$~MFLOPs & $244$~MFLOPs & $5.9$~MB\\\hline
		\textbf{MCAO} & $63.5$~MFLOPs & $33.5$~MFLOPs & $0.78$~MB \\\hline
	\end{tabular}
	\caption{Performance evaluation of augmented FEWHA versus classical FEWHA.}
	\label{tab:FEWHAFLOPsELT}
\end{table} 

\subsection{Performance on CPU} 
For ELT-sized test settings FEWHA performs better on CPUs than on GPUs. This is mainly because of its efficient matrix-free implementation. The bottleneck is not the computational throughput but the memory latency. For a detailed study we refer to \cite{stadler2020}. That is why we focus here on the implementation of augmented FEWHA on CPU. In particular, we run the parallel implementation of our method on one compute node of the high performance cluster of the Radon Institute for Computational and Applied Mathematics in Linz, called Radon1, that has two 8-core Intel Haswell processors (Xeon 403 E5-2630v3, 2.4Ghz) and 128 GB of memory. For our numerical simulations we utilize $12$ cores. We reuse the parallel implementation of FEWHA from \cite{stadler2020}, but change the PCG method to augmented PCG. 

The last two columns of Table \ref{tab:FEWHARuntimeELT} refer to the runtime of the certain test configurations for FEWHA and its augmented version. We simulate $1000$ time steps and take the average runtime. We gain a speed-up of augmented FEWHA compared to the classical method between $1.63$ and $1.8$. Of course the speed-up of the MCAO 3-layer setting is less compared to the other two settings, because we use a smaller number of iterations, and thus the computations outside the PCG algorithm have a bigger influence on the overall performance. This test case is particularly important as it is related to MAORY, which is an adaptive optics module for the ELT operating in MCAO. The real-time requirement of MAORY is $2$~ms for each reconstruction. To the best of our knowledge the augmented FEWHA is the first method that is able to solve the 3-layer MCAO test setting within this real-time requirement.

\begin{table}[ht]
	\center
	\footnotesize
	\renewcommand{\arraystretch}{1.3}
	\begin{tabular}{|c|c|c|c|}
		\hline
		\textbf{Test setting} & \textbf{Runtime FEWHA} & \textbf{Runtime aug.~FEWHA} & \textbf{Speed-up}\\\hline
		\textbf{LTAO}  & $6$~ms & $3.5$~ms & $1.7$\\\hline
		\textbf{MOAO} & $8$~ms & $4.4$~ms & $1.8$\\\hline
		\textbf{MCAO} & $3.1$~ms & $1.9$~ms & $1.63$ \\\hline
	\end{tabular}
	\caption{Runtime of augmented FEWHA versus classical FEWHA on Radon1.}
	\label{tab:FEWHARuntimeELT}
\end{table}  

\section{Conclusion}
In this paper, we continued the work of \cite{Yu14, HeYu13,YuHeRa13, YuHeRa13b, stadler2020} on the iterative solver \nolinebreak FEWHA by considering a Krylov subspace recycling method. This method enables to use information from previous time steps to speed up the convergence of the iterative PCG method. The inner products and matrix-vector products within FEWHA are well parallelizable, however, the number of PCG iterations are a crucial indicator for the real-time performance. Numerical simulations for ELT-sized problems showed that we are able to decrease the number of iterations to $50\%$ with augmented FEWHA. Assuming half the PCG iterations, we provided a detailed study on the computational performance, i.e. FLOPs, memory usage and runtime on CPU. The only drawback of augmented FEWHA is the additional used memory to store information from previous time steps. However, this is almost negligible compared to the overall memory usage. Most importantly, we showed that with augmented FEWHA we are able to meet the real-time requirements of MAORY, an adaptive optics module for the ELT operating in MCAO. To the best of our knowledge, augmented FEWHA is the first method that fulfills this requirement.

\section{Acknowledgments}
The project has received funding by the European Union’s Horizon 2020 research and innovation programme under the Marie Sk\l odowska-Curie Grant Agreement No.~765374 and the Austrian Science Fund (FWF) F6805-N36 (Tomography in Astronomy).

\bibliography{literature}
\bibliographystyle{siam}
\end{document}